\documentclass[sigplan,10pt]{acmart}

\setcopyright{none}
\renewcommand\footnotetextcopyrightpermission[1]{}
\acmYear{2026}
\copyrightyear{2026}
\acmConference[ATC '26]{The 2026 ACM SIGOPS Annual Technical Conference}{November 16--18, 2026}{Hong Kong}
\acmBooktitle{The 2026 ACM SIGOPS Annual Technical Conference (ATC '26), November 16--18, 2026, Hong Kong}
\acmDOI{}
\acmISBN{}

\usepackage{booktabs}
\usepackage{amsmath}
\usepackage{array}
\usepackage{algorithm}
\usepackage{algpseudocode}
\usepackage{subcaption}
\usepackage{xspace}

\newif\ifdraft
\draftfalse

\newcommand{\sys}{BiDiRL}
\title{\sys: Bidirectional Resource Scheduling for Disaggregated and Asynchronous RL Post-Training}

\author{Zhiqiang Tan}
\email{25b951103@stu.hit.edu.cn}
\affiliation{%
  \institution{School of Computer Science and Technology, Harbin Institute of Technology, Shenzhen}
  \city{Shenzhen}
  \country{China}
}

\author{Maoxin Wang}
\email{25s151169@stu.hit.edu.cn}
\affiliation{%
  \institution{School of Computer Science and Technology, Harbin Institute of Technology, Shenzhen}
  \city{Shenzhen}
  \country{China}
}

\author{Sijie Wang}
\email{25b951105@stu.hit.edu.cn}
\affiliation{%
  \institution{School of Computer Science and Technology, Harbin Institute of Technology, Shenzhen}
  \city{Shenzhen}
  \country{China}
}

\author{Yiming Yin}
\email{yyin464@connect.hkust-gz.edu.cn}
\affiliation{%
  \institution{Data Science and Analytics Thrust, The Hong Kong University of Science and Technology (Guangzhou)}
  \city{Guangzhou}
  \country{China}
}

\author{Qiang Wang}
\email{qiang.wang@hit.edu.cn}
\affiliation{%
  \institution{School of Computer Science and Technology, Harbin Institute of Technology, Shenzhen}
  \city{Shenzhen}
  \country{China}
}

\author{Xiaowen Chu}
\email{xwchu@ust.hk}
\affiliation{%
  \institution{Data Science and Analytics Thrust, The Hong Kong University of Science and Technology (Guangzhou)}
  \city{Guangzhou}
  \country{China}
}

\author{Shaohuai Shi}
\authornote{Corresponding author.}
\email{shaohuais@hit.edu.cn}
\affiliation{%
  \institution{School of Computer Science and Technology, Harbin Institute of Technology, Shenzhen}
  \city{Shenzhen}
  \country{China}
}

\begin{document}
\fancypagestyle{plain}{%
  \fancyhf{}%
  \fancyfoot[C]{\footnotesize\thepage}%
  \renewcommand{\headrulewidth}{0pt}%
  \renewcommand{\footrulewidth}{0pt}%
}
\pagestyle{plain}
\addtolength{\textheight}{-5pt}

\begin{abstract}
It is well established that the reasoning capabilities of large language models (LLMs) can be improved by applying reinforcement learning (RL) in a post-training stage. In a standard RL iteration, the current model (the policy) generates experience through rollouts, and the resulting data is then used to update the policy during training. High-performance RL frameworks such as StreamRL and AReaL employ a disaggregated architecture and asynchronous rollouts to better exploit both rollout and training resources, thereby increasing overall system throughput. 
Nonetheless, across varying RL setups (e.g., hardware configurations, model scales, staleness levels, and hyperparameters) and under changing workloads, it remains common for both rollout and training resources to experience idle periods.

In this paper, we present BiDiRL, a hybrid time-space multiplexing architecture for asynchronous, disaggregated RL designed to reduce resource idleness. 
First, we develop a hot-switch runtime that enables rapid switching between rollout and training resources with negligible overhead. 
Second, we propose a static, scheduling-aware planner based on time-performance modeling that chooses a hot-switch-friendly resource partition, so that rollout and training durations are roughly balanced at a coarse level. 
Third, at execution time, we introduce a bidirectional scheduler that further exploits runtime bubbles through fine-grained resource switching, allowing the bottleneck stage to temporarily borrow idle resources from the other pool. 
Across a wide range of workloads, datasets, and models on two 32-GPU testbeds, 
BiDiRL increases RL training throughput by up to \(1.94\times\) compared with 
state-of-the-art RL systems—including veRL, AReaL, and ROLL—without affecting convergence behavior.
\end{abstract}

\maketitle

\section{Introduction}
\label{sec:introduction}

Reinforcement learning (RL) post-training has become a key technique for improving the reasoning,
coding, and tool-use capabilities of
large language models (LLMs)~\cite{deepseek-aiDeepSeekR1IncentivizingReasoning2025,shaoDeepSeekMathPushingLimits2024,NEURIPS2022_b1efde53,liu-etal-2025-epo,wu2024framework}.
An LLM RL job mainly consists of two stages.
In the rollout stage, the actor model takes a batch of prompts and generates responses
using an inference engine such as vLLM~\cite{kwonEfficientMemoryManagement2023}.
We refer to the workers that execute this stage as \emph{rollouters}.
In the training stage, the system scores generated responses,
computes policy losses under objectives such as group relative policy optimization (GRPO)~\cite{shaoDeepSeekMathPushingLimits2024},
and updates the actor model using a training engine such as PyTorch FSDP~\cite{zhaoPyTorchFSDPExperiences2023}.
We refer to the workers that execute this stage as \emph{trainers}.
After each update, the latest actor weights are transferred to rollouters for subsequent generation.

% Rollout and training exhibit distinct execution characteristics~\cite{zhongStreamRLScalableHeterogeneous}.
% Rollout first processes prompt tokens through prefill and then generates response tokens
% through autoregressive decoding~\cite{zhongDistServeDisaggregatingPrefill}.
% Its execution time is therefore sensitive to response length,
% because a few long-tail sequences can dominate the overall rollout time~\cite{zhangSortedRLAcceleratingRL2026,shengLaminarScalableAsynchronous2026}.
% Training consumes completed responses through batched model execution,
% including policy log-probability recomputation,
% reference-policy evaluation when required, and actor updates~\cite{shaoDeepSeekMathPushingLimits2024}.
% % These two stages exhibit distinct execution characteristics~\cite{zhongStreamRLScalableHeterogeneous}.
% As shown in Figure~\ref{fig:intro-workload-scaling} and Figure~\ref{fig:intro-resource-scaling},
% rollout is more sensitive to longer responses and usually obtains smaller gains from additional resources than training,
% while training can expose denser batched computation over tokens~\cite{fuAReaLLargeScaleAsynchronous2025,zhongStreamRLScalableHeterogeneous}.

Existing LLM RL systems mainly organize rollout and training with either colocated or disaggregated architectures.
A colocated architecture improves device utilization by time-sharing a single resource pool
between rollout and training, as in veRL~\cite{shengHybridFlowFlexibleEfficient2025} and DistFlow~\cite{wang2026distflow}.
Since the two stages occupy the same devices at different times,
colocation avoids dedicating resources to a stage that is temporarily inactive.
However, this utilization benefit comes at the cost of resource coupling:
rollout and training are constrained to the same resource pool despite their distinct scaling behavior~\cite{zhongStreamRLScalableHeterogeneous}.
Recent systems revisit disaggregated execution by combining separate rollout and training resource pools with asynchronous or off-policy training~\cite{fuAReaLLargeScaleAsynchronous2025,wangReinforcementLearningOptimization2025,zhongStreamRLScalableHeterogeneous}.
By allowing trainers to consume samples generated by recent actor versions,
these systems overlap rollout and training and improve overall throughput.

\begin{table}[!t]
  \centering
  \caption{Comparison with representative RL systems.}
  \label{tab:system-comparison}
  \scriptsize
  \setlength{\tabcolsep}{4.0pt}
  \renewcommand{\arraystretch}{1.08}
  \begin{tabular}{@{}lcccccc@{}}
    \toprule
    Feature & \textbf{BiDiRL} & veRL & RLBoost & StreamRL & AReaL & ROLL \\
    \midrule
    Disaggregated & \(\surd\) & \(\surd\) & \(\surd\) & \(\surd\) & \(\surd\) & \(\surd\) \\
    Off-policy & \(\surd\) & \(\surd\) & \(\surd\) & \(\surd\) & \(\surd\) & \(\surd\) \\
    Fractional staleness & \(\surd\) & \(\surd\) & \(\times\) & \(\times\) & \(\times\) & \(\times\) \\
    Static planning & \(\surd\) & \(p\) & \(\times\) & \(p\) & \(\times\) & \(\times\) \\
    Same-budget borrow & \(\surd\) & \(\times\) & \(\times\) & \(\times\) & \(\times\) & \(\times\) \\
    Elastic rollout & \(\surd\) & \(\times\) & \(\surd\) & \(\surd\) & \(\times\) & \(\times\) \\
    Elastic training & \(\surd\) & \(\times\) & \(\times\) & \(\times\) & \(\times\) & \(\times\) \\
    \bottomrule
  \end{tabular}
  % \vspace{0.25em}
  \begin{minipage}{0.98\columnwidth}
    \scriptsize \(\surd\), \(\times\), and \(p\) denote supported, unsupported, and partially supported, respectively.
    Fractional staleness denotes staleness bounds \(0<s<1\).
    Same-budget elasticity means temporarily expanding rollout or training with idle resources from the other committed pool without adding GPUs.
  \end{minipage}
\end{table}

Nevertheless, overlap alone does not remove idle resource bubbles left by fixed disaggregated allocation.
As detailed in Section~\ref{sec:bg-motivation}, these bubbles have two sources.
Some are predictable before execution: a fixed rollout-training partition can create persistent rate mismatch.
Others arise during execution: response-length dynamics~\cite{deepseek-aiDeepSeekR1IncentivizingReasoning2025,liuDAPOImprovingMultiStep,tanArlEfficientAgentic2026,tan2026orchestrrl}, constraints imposed by model parallelism strategies,
and staleness constraints can leave idle windows even under a well-chosen partition.
We refer to these two cases as \emph{structural bubbles} and \emph{residual bubbles}, respectively.
Existing systems attack these bubbles from complementary directions, as summarized in Table~\ref{tab:system-comparison}.
Asynchronous and off-policy systems increase overlap, static planners improve the initial partition,
and rollout-side schedulers or external resource harvesting reduce generation-side stalls~\cite{zhongStreamRLScalableHeterogeneous,wuWeave2026,fuAReaLLargeScaleAsynchronous2025,wangReinforcementLearningOptimization2025,wuRLBoostHarvestingPreemptible2025}.
These approaches leave a resource-orchestration challenge.
Static planning can reduce structural bubbles before execution, but it cannot reclaim residual bubbles caused by runtime variation and staleness.
Rollout-side scheduling or offloading helps when rollout is the bottleneck,
but it does not address trainer-heavy or two-sided idle windows within the same committed budget.
Thus, disaggregated RL needs a mechanism that chooses a good initial partition and,
during execution, reuses whichever committed resource pool becomes temporarily idle without violating staleness constraints.

To address this challenge, we develop BiDiRL, a hybrid time-space multiplexing system for disaggregated and asynchronous RL post-training.
BiDiRL keeps rollout and training on separate committed resource pools to preserve disaggregated overlap,
but allows either pool to temporarily execute the auxiliary role of the other stage when doing so is predicted to reduce idle time. To achieve this goal, we first design a hot-switch runtime (\S\ref{sec:runtime-support}) to make rollouter and trainer roles exchangeable on their resource pools with negligible overhead.
Then, before execution, we design a scheduling-aware static planner (\S\ref{sec:static-planning}) to select a rate-balanced resource partition and construct a hot-switch-compatible resource envelope with time performance modeling.
Finally, during execution, we design a bidirectional scheduler (\S\ref{sec:bidirectional-scheduling}) to admit borrowing resources from different pools when the predicted benefit exceeds the measured switching overhead, and it also decides how much workload should be assigned to primary and auxiliary workers with our real-time online profiling (\S\ref{sec:online-profiler}).
These mechanisms allow BiDiRL to harvest any possible idle resource window introduced by disaggregation during RL post-training.

We evaluate BiDiRL against state-of-the-art open-source RL systems,
including veRL~\cite{shengHybridFlowFlexibleEfficient2025}, AReaL~\cite{fuAReaLLargeScaleAsynchronous2025}, and ROLL~\cite{wangReinforcementLearningOptimization2025}.
The evaluation covers diverse workloads, staleness bounds, resource budgets, model sizes, and two 32-GPU platforms (NVIDIA A6000 and H100).
Across all the evaluated cases, BiDiRL runs faster by $1.05\times$--$1.94\times$ over veRL, AReaL, and ROLL.
It improves effective training throughput by \(1.27\times\)--\(1.68\times\) across the 16-GPU A6000 testbed and reaches \(1.94\times\) with 32 GPUs.
Similarly, on the H100 testbed, \sys{} achieves \(1.23\times\)--\(1.53\times\) improvement.
Ablations show that our dynamic bidirectional scheduler improves over no borrowing by \(1.12\times\)--\(1.71\times\)
and over opportunistic borrowing by \(1.02\times\)--\(1.31\times\).
% Training-behavior experiments show reward trajectories aligned with veRL under the same staleness settings.
In summary, we make the following contributions:
\begin{itemize}
\item We identify idle resource bubbles in disaggregated LLM RL under staleness bounds as a two-timescale problem:
structural bubbles arise from static partition mismatch, while residual bubbles come from workload dynamics,
model-parallelism constraints, and staleness constraints.

\item We design a hot-switch runtime to enable rollout and training to be switchable in runtime with negligible overhead.

\item We design a scheduling-aware static planner that selects the initial rollout-training resource partitioning
and constructs a hot-switch-compatible resource envelope, exposing lightweight prediction interfaces used by runtime scheduling.

\item We propose a bidirectional scheduler that dynamically borrows idle rollout or training resources for the bottleneck stage,
deciding both when to borrow and how much workload to assign using a benefit-over-overhead admission rule to minimize resource idleness.

\end{itemize}

\section{Background and Motivation}
\label{sec:background}

\subsection{LLM RL Workflow}
\label{sec:bg-rl-workflow}

In LLM RL post-training, the actor is the model being optimized.
Given a batch of prompts, the actor generates responses autoregressively and is trained to maximize their expected reward.
In GRPO-style workloads, multiple responses generated from the same prompt constitute a prompt-level rollout group,
while each completed response is treated as a response-level training sample~\cite{deepseek-aiDeepSeekR1IncentivizingReasoning2025}.
Modern rollout engines, such as vLLM~\cite{kwonEfficientMemoryManagement2023} and SGLang~\cite{zhengSGLangEfficientExecution},
combine model parallelism with data parallelism to create inference replicas.

The training stage consumes completed response-level samples.
These operations are commonly executed by training engines based on PyTorch FSDP~\cite{zhaoPyTorchFSDPExperiences2023}
or Megatron-LM~\cite{shoeybiMegatronLMTrainingMultiBillion2020}.
To reduce memory usage, training engines typically partition a batch into micro-batches,
process them sequentially, and accumulate gradients across micro-batches before applying the update.
In BiDiRL, we use the term \emph{chunk} to denote the executable scheduling unit induced by this micro-batching.
Its size is determined by the per-device micro-batch size and the data-parallel degree,
and it serves as the unit used by Trainer-on-RollPoll scheduling (Section~\ref{sec:bidirectional-scheduling}).
This decomposition provides the basis for modeling trainer execution time.

\subsection{Staleness-Bounded Disaggregated RL}
\label{sec:bg-disaggregated}

Disaggregated RL systems map rollout and training to separate committed resource pools. 
This execution model allows rollouters and trainers to use specialized engines and overlap sample generation with model update.
Asynchronous execution is governed by a staleness policy \(s\).
At the sample level, if a sample is generated by actor version \(v_g\) and consumed when the trainer is updating actor version \(v_t\),
its actor-version distance is \(\Delta v=v_t-v_g\)~\cite{fuAReaLLargeScaleAsynchronous2025}.
A sample is valid only if it satisfies the configured staleness policy.
At the batch level, some systems use a floating-point policy \(s\),
where \(s<1\) caps the stale-sample share and \(s\ge1\) removes the batch-level freshness requirement~\cite{shengHybridFlowFlexibleEfficient2025}.
Thus, \(s=0\) indicates strict on-policy execution, where all samples must be generated by the latest actor version.
When \(0<s<1\), trainers may consume a mixture of new and recent valid stale samples, but each batch still requires enough new samples.
When \(s\ge1\), an entire batch may be formed from recent valid samples.

Modern disaggregated systems often use partial rollout~\cite{zhangSortedRLAcceleratingRL2026} to avoid blocking on long responses.
When trainers reach a synchronization point, unfinished generations can be interrupted and carried over to the next window.
With partial rollout, the steady-state execution can be viewed from the trainer side as two phases:
waiting until enough valid rollout groups are available and then consuming the resulting batch.
Ignoring the initial buffer fill and the final drain, and folding weight synchronization into trainer consumption unless stated otherwise,
we approximate the duration of one training window as
\begin{equation}
T_{\mathrm{step}} \approx T_{\mathrm{wait}} + T_{\mathrm{consume}},
\end{equation}
where \(T_{\mathrm{wait}}\) is the time until the rollout buffer contains enough valid rollout groups under the staleness bound,
and \(T_{\mathrm{consume}}\) is the time for trainers to consume the batch.
Without partial rollout, the current window remains tied to the slowest response in the rollout batch,
making the step time closer to the maximum of rollout completion time and trainer consumption time.
Partial rollout removes this tail from the current window, but any residual shortage of valid samples appears as \(T_{\mathrm{wait}}\) in the next window.

\subsection{Motivation: Resource Bubbles}
\label{sec:bg-motivation}

\begin{figure}[t]
  \centering
  \captionsetup[subfigure]{font=small,justification=centering,singlelinecheck=false}
  \begin{subfigure}[t]{0.48\linewidth}
    \centering
    \includegraphics[width=\linewidth]{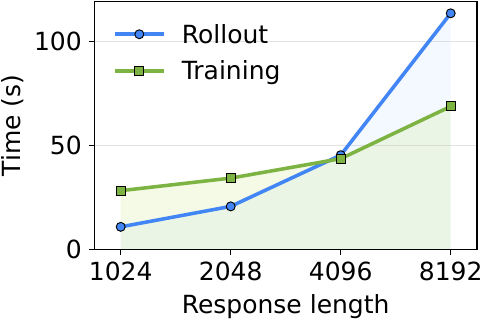}
    \caption{Response length scaling.}
    \Description{Line chart comparing rollout and training profile time as generated response length increases.}
    \label{fig:intro-workload-scaling}
  \end{subfigure}
  \hfill
  \begin{subfigure}[t]{0.48\linewidth}
    \centering
    \includegraphics[width=\linewidth]{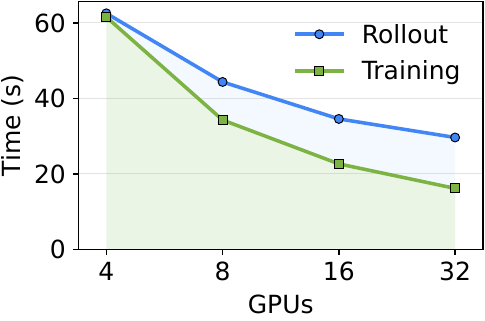}
    \caption{Resource scaling.}
    \Description{Line chart comparing rollout and training profile time as the number of GPUs increases at fixed response length.}
    \label{fig:intro-resource-scaling}
  \end{subfigure}
  \vspace{0.5ex}

  \begin{subfigure}[t]{0.48\linewidth}
    \centering
    \includegraphics[width=\linewidth]{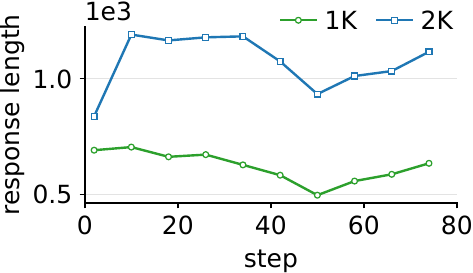}
    \caption{Response length dynamics.}
    \Description{Line chart showing that generated response length varies over RL training steps.}
    \label{fig:intro-response-dynamics}
  \end{subfigure}
  \hfill
  \begin{subfigure}[t]{0.48\linewidth}
    \centering
    \includegraphics[width=\linewidth]{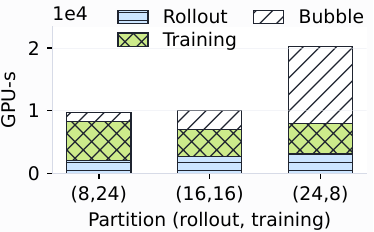}
    \caption{Bubbles with partitions.}
    \Description{Time breakdown showing waiting time caused by a static resource partition mismatch between rollout and training.}
    \label{fig:intro-static-split-bubbles}
  \end{subfigure}
  \caption{Profiling evidence for resource bubbles in disaggregated LLM RL.
  Rollout and training exhibit different scaling trends~\cite{zhongStreamRLScalableHeterogeneous} (top row).
  Static partitions still leave resource bubbles (bottom row).}
  \Description{Four panels showing rollout-training sensitivity to response length, sensitivity to GPU resources, changing response length over RL training steps, and static resource partition waiting time.}
  \label{fig:intro-rollout-trainer-scaling}
\end{figure}

Rollout and training exhibit distinct execution characteristics~\cite{zhongStreamRLScalableHeterogeneous}.
Rollout first processes prompt tokens through prefill and then generates response tokens through autoregressive decoding~\cite{zhongDistServeDisaggregatingPrefill}.
Its execution time is sensitive to response length because a few long-tail sequences can dominate the overall rollout time~\cite{zhangSortedRLAcceleratingRL2026,shengLaminarScalableAsynchronous2026,gao2026rollpacker}.
Training consumes completed responses through batched model execution,
including policy log-probability recomputation, reference-policy evaluation when required, and actor updates~\cite{shaoDeepSeekMathPushingLimits2024}.
As shown in Figure~\ref{fig:intro-workload-scaling} and Figure~\ref{fig:intro-resource-scaling},
rollout is more sensitive to longer responses and usually obtains smaller gains from additional resources than training,
while training can expose denser batched computation over tokens~\cite{fuAReaLLargeScaleAsynchronous2025,zhongStreamRLScalableHeterogeneous}.

These scaling differences create bubbles at two time scales.
Before execution, the system must partition a fixed resource budget into committed rollout and training pools.
This partitioning problem appears in existing disaggregated systems~\cite{zhongStreamRLScalableHeterogeneous,huOpenRLHFEasytouseScalable,yao2023deepspeed,shengLaminarScalableAsynchronous2026}.
A poor resource partition creates a persistent rate mismatch, where one stage repeatedly waits because the other stage is slower.
We refer to the predictable idle windows caused by static resource partition mismatch as \emph{structural bubbles}.
During execution, even a well-tuned initial partition cannot eliminate all idle time.
Response lengths may shift as the policy evolves during RL training (Figure~\ref{fig:intro-response-dynamics})~\cite{deepseek-aiDeepSeekR1IncentivizingReasoning2025,liuDAPOImprovingMultiStep,tanArlEfficientAgentic2026,tan2026orchestrrl}.
Feasible resource allocations are also constrained by model parallelism strategies, which limits how finely the partitioning can be tuned.
A strict staleness bound can further prevent rollouters from running beyond the permitted run-ahead window,
or force trainers to wait until enough new rollout samples become available.
We refer to the remaining idle windows caused by workload variation, model parallelism constraints, and staleness constraints as \emph{residual bubbles}.
As shown in Figure~\ref{fig:intro-static-split-bubbles}, such bubbles remain even under different static resource partitioning.

\begin{figure}[t]
  \centering
  \captionsetup[subfigure]{font=small,justification=centering,singlelinecheck=false}
  \begin{subfigure}[t]{0.32\linewidth}
    \centering
    \includegraphics[width=\linewidth]{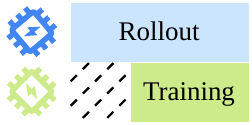}
    \caption{Training idle}
    \Description{Timeline showing asynchronous disaggregated execution where rollout takes longer than training and trainers wait for rollout samples.}
    \label{fig:motivation-rollout-bound}
  \end{subfigure}
  \hfill
  \begin{subfigure}[t]{0.32\linewidth}
    \centering
    \includegraphics[width=\linewidth]{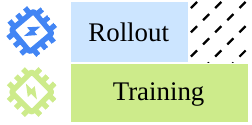}
    \caption{Rollout idle}
    \Description{Timeline showing asynchronous disaggregated execution where training takes longer than rollout and rollouters wait for trainers to finish.}
    \label{fig:motivation-trainer-bound}
  \end{subfigure}
  \hfill
  \begin{subfigure}[t]{0.32\linewidth}
    \centering
    \includegraphics[width=\linewidth]{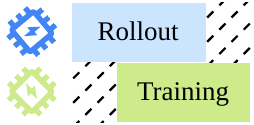}
    \caption{Two-sided idle}
    \Description{Timeline showing asynchronous disaggregated execution where trainers wait for enough new rollout samples and fast rollouters leave tail bubbles under the staleness bound.}
    \label{fig:motivation-bidirectional-bound}
  \end{subfigure}
  \caption{Motivation for bidirectional scheduling in disaggregated LLM RL.
  With bounded off-policy execution, rollout and training can overlap but bubbles remain when (a) rollout is slower, (b) training is slower, or (c) the staleness bound forces trainers to wait for fresh samples after rollouters run ahead.}
  \Description{Three timeline diagrams comparing asynchronous off-policy disaggregation cases with rollout-bound, trainer-bound, and staleness-bound residual idle bubbles.}
  \label{fig:motivation-bubbles}
\end{figure}

The side on which residual bubbles appear depends on the workload, the staleness bound, and the resource partition.
Long-reasoning and multi-turn tasks tend to be rollout-heavy because they generate longer responses (Figure~\ref{fig:motivation-rollout-bound})~\cite{wuAgenticReasoningStreamlined,gao2025rollart}.
In contrast, short-output workloads may expose a training-side bottleneck (Figure~\ref{fig:motivation-trainer-bound}),
especially for multimodal RL tasks where responses are short but training still processes full multimodal inputs~\cite{fengVideoR1ReinforcingVideo2025,nvidiaCosmosWorldFoundation2025,wang2025alpamayo}.
With a strict staleness bound, bubbles may also appear on both sides:
trainers may wait for enough new rollout samples, while rollouters stop after exhausting the permitted run-ahead window (Figure~\ref{fig:motivation-bidirectional-bound}).
These cases show that idle resources in disaggregated RL are not always one-sided.
Either the rollout pool or the training pool may become temporarily idle.

This observation motivates the two-timescale design used by BiDiRL.
Static planning selects a parallelism-compatible partition that reduces structural bubbles,
while runtime bidirectional scheduling reclaims the residual bubbles that fixed allocation cannot fully eliminate.

\section{\sys{}: System Design}
\label{sec:overview}
\begin{figure}[t]
  \centering
  \includegraphics[width=\columnwidth]{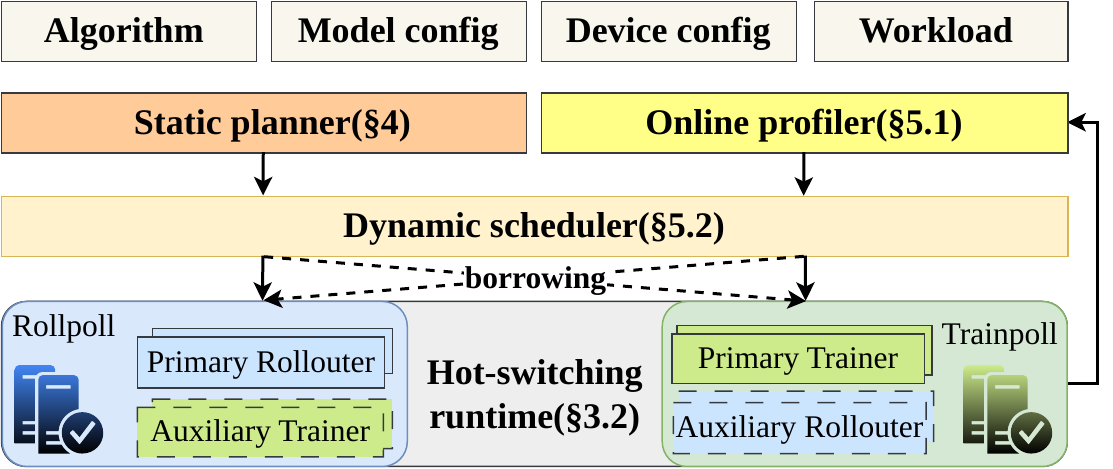}
  \caption{The system overview of \sys{}.}
  \label{fig:BiDiRL-architecture}
\end{figure}

\subsection{System Overview}
Figure~\ref{fig:BiDiRL-architecture} shows the architecture of \sys{}.
% BiDiRL runs a staleness-bounded disaggregated RL job on two committed resource pools.
For a given GPU cluster for RL post-training, \sys{} partitions the GPU resources into two pools: a rollout pool (RollPoll) for rollouters and a training pool (TrainPoll) for trainers. 

The key idea of \sys{} is to maximally reduce the bubble time under the disaggregated architecture for different RL training scenarios by dynamically enabling trainers to run on RollPoll and rollouters to run on TrainPoll in runtime. Thus, we develop a hot-switch runtime (\S\ref{sec:runtime-support}) that supports lightweight context switching, making the associated overhead negligible (\S\ref{sec:eval-ablation}). 

Then, \sys{} introduces a static scheduling-aware planner (\S\ref{sec:static-planning}) to coarsely assign a proper number of GPUs as RollPoll and TrainPoll so that the rollout time and training time are comparable and both resources are switchable in runtime. The key idea of the static planner is to model time performance (including compute and communication) of different inference and training tasks under different GPU resources with a constraint of allowing hot switching between rollouters and trainers, so that we can assign a proper group of GPUs for RollPoll and TrainPoll.

After the static planning of resources, \sys{} launches the RL post-training job by placing rollouters in RollPoll and trainers in TrainPoll. During RL execution, \sys{} exploits our proposed bidirectional scheduler (\S\ref{sec:bidirectional-scheduling}) to dynamically determine when rollouters should borrow resources from TrainPoll and when trainers should borrow resources from RollPoll to minimize the iteration time. Using RollPoll for trainers (Trainer-on-RollPoll) and TrainPoll for rollouters (Rollouter-on-TrainPoll) in runtime requires dynamically switching the context of rollout and training. For example, in Trainer-on-RollPoll, a training instance with the newest weight parameters should be launched and collaborate with the existing trainers on TrainPoll. To accurately ensure that bidirectional resource switching can always bring performance benefits, \sys{} introduces a lightweight online profiler (\S\ref{sec:runtime-support}) that measures system time performance (e.g., rollout time, training time, weight synchronization time, etc.). Based on the profiling information and workload models, \sys{} formulates each borrowing opportunity as an admission-control and workload-splitting problem.
A borrowed window is admitted only when the predicted benefit outweighs the measured hot-switch overhead, and the assigned workload is sized according to the predicted completion time of the primary and auxiliary workers (\S\ref{sec:bidirectional-scheduling}).

\begin{figure}[t]
  \centering
  \begin{subfigure}[t]{0.96\linewidth}
    \centering
    \includegraphics[width=\linewidth]{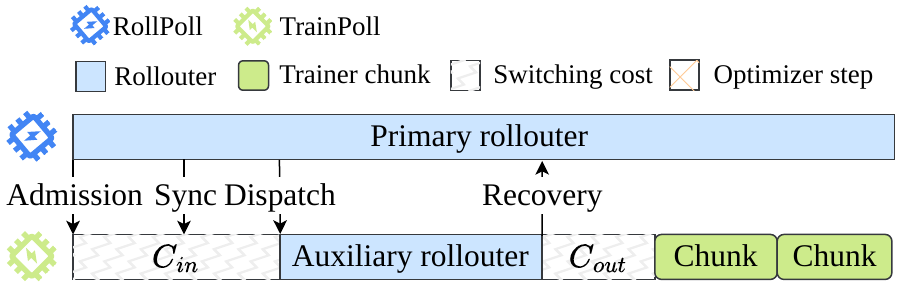}
    \caption{Rollouter-on-TrainPoll: rollouters borrow the training pool.}
    \Description{Scheduling diagram showing rollout work borrowing idle trainer-pool resources.}
    \label{fig:rollout-on-training}
  \end{subfigure}
  \vspace{0.75em}
  \begin{subfigure}[t]{0.96\linewidth}
    \centering
    \includegraphics[width=\linewidth]{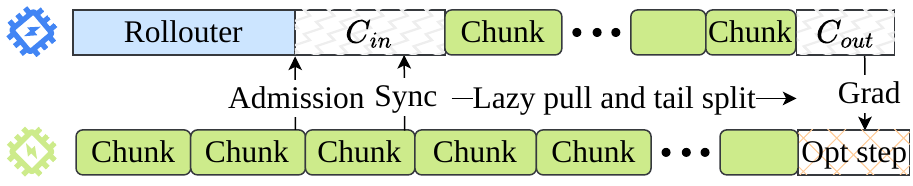}
    \caption{Trainer-on-RollPoll: trainers borrow the rollout pool.}
    \Description{Scheduling diagram showing trainer work borrowing idle rollout-pool resources.}
    \label{fig:trainer-on-rollout}
  \end{subfigure}
  \caption{Bidirectional scheduling. The two directions share an overhead-aware admission interface but use different work units and recovery rules.}
  \label{fig:bidirectional}
\end{figure}

\subsection{Hot-Switching Runtime}
\label{sec:runtime-support}

% \paragraph{Hot switching.}
Hot switching for rollouters and trainers is the runtime mechanism that makes the planner-selected resource envelope executable (details in \S\ref{sec:static-planning}) and switchable (details in \S\ref{sec:bidirectional-scheduling}).
At startup, auxiliary workers are initialized and then kept passive.
Primary workers own their pools by default.
% A borrowed window grants the auxiliary workers a temporary lease on the pool.
As shown in Figure~\ref{fig:bidirectional},
when the scheduler admits borrowing, the runtime releases the primary workers and wakes the auxiliary workers.
When a primary worker needs to resume or the auxiliary workers finish their assigned tasks,
the runtime revokes the auxiliary lease.
The auxiliary workers then stop accepting new tasks, return recoverable pending tasks,
and yield the pool back to the primary workers.
This preempt-and-yield rule ensures that a pool is never used by rollout and training workers at the same time.
It also gives the scheduler flexibility to switch without restarting the rollout or training instance.

For Rollouter-on-TrainPoll (Figure~\ref{fig:rollout-on-training}), \(C_{\mathrm{in}}\) indicates the time for switching the training pool from primary trainers to auxiliary rollouters,
which involves offloading trainer states, loading rollouter states, and synchronizing with primary rollouters when needed.
\(C_{\mathrm{out}}\) indicates the time for switching back to primary trainers,
including offloading rollouter states and reloading trainer states.
After borrowing, complete rollout samples are committed normally, and incomplete rollout samples are returned as partial rollout state for primary rollouters to resume.

For Trainer-on-RollPoll (Figure~\ref{fig:trainer-on-rollout}), \(C_{\mathrm{in}}\) indicates the time for switching the rollout pool from primary rollouters to auxiliary trainers,
and \(C_{\mathrm{out}}\) indicates the time for switching back to primary rollouters,
which includes gradient synchronization if the borrowed window covers actor update.
% When the borrowed window includes actor update,
% gradients computed by the auxiliary trainers are merged back into the primary trainer before the update is applied.
When the borrowed window ends, trainer chunks that have not started execution can return to the pending queue,
while running chunks finish and merge in original order; Section~\ref{para:chunk-pipeline}
summarizes the chunk state classes used by these recovery rules.

\section{Scheduling-Aware Static Planning}
\label{sec:static-planning}

Static planning runs once before RL training starts and returns a resource envelope
\begin{equation}
\mathcal{E}=(g_r,g_t,\rho_r,\rho_t,\mathcal{M}_r,\mathcal{M}_t),
\end{equation}
where \(g_r\) and \(g_t\) are the number of GPUs assigned to RollPoll and TrainPoll,
\(\rho_r\) and \(\rho_t\) are the replica layouts used by rollout and training.
\(\mathcal{M}_r,\mathcal{M}_t\) are the stage-time models for rollout and training, respectively. Due to the page limit, we leave the details of modeling to Appendix~\ref{app:planner-models} and they can be accurately verified according to our experimental results (\S\ref{sec:eval-static}).

This envelope $\mathcal{E}$ is constructed to be both hot-switch compatible and throughput-aware.

\subsection{Hot-switch Compatible Model Layouts}
The planner first determines the minimal feasible model layout for each stage.
A model layout specifies the number of devices occupied by one rollout or trainer replica.
BiDiRL follows a minimal-replica principle: once a stage replica can satisfy the memory and execution constraints,
using a larger replica usually reduces the data-parallel degree and may introduce additional communication overheads.
The selected layouts also define the granularity of hot switching.
For a candidate resource partition to be legal,
each resource pool must be divisible into an integer number of replicas for every role it may host.
% For any resource pool that may host a stage, either as the primary role or as the auxiliary role,
% the number of devices in the pool must contain an integer number of that stage's replicas.
This constraint allows the auxiliary role to be activated without process restarts or expensive resharding.
Primary and auxiliary workers of the same stage also use the same model layout,
so model weights and gradients can be transferred without layout resharding.

This constraint in our \sys{} distinguishes it from the existing planner that only balances stage throughput~\cite{shengHybridFlowFlexibleEfficient2025,zhongStreamRLScalableHeterogeneous,meiReaLEfficientRLHFa}.
A resource partition with good predicted throughput can still be unusable if it cannot enable hot switching.
BiDiRL therefore searches only over partitions that satisfy both throughput and compatibility requirements as follows.

\subsection{Execution Time Models and Resource Allocation}
Given the compatible model layouts, the planner enumerates legal rollout-training partitions under the total resource budget.
For each candidate partition \((g_r,g_t)\), it derives the rollout and trainer replica counts
\(d_r=\Call{ReplicaCount}{g_r,\rho_r}\) and \(d_t=\Call{ReplicaCount}{g_t,\rho_t}\).
We write \(\mathcal{M}_r(n,d)\) as the predicted time for \(d\) rollout replicas to produce \(n\) prompt-level rollout groups,
and \(\mathcal{M}_t(U,d)\) as the predicted time for \(d\) trainer replicas to consume a stage-ordered trainer workload \(U\).
For the representative batch in the workload statistics \(W\), let \(B_p\) be the number of prompt-level rollout groups
and \(U_B\) be the corresponding trainer workload.
The planner computes
\begin{equation}
\lambda_r=\frac{B_p}{\mathcal{M}_r(B_p,d_r)},
\qquad
\lambda_t=\frac{B_p}{\mathcal{M}_t(U_B,d_t)}.
\end{equation}
The selected partition maximizes the steady-state pipeline rate
\begin{equation}
\lambda_{\mathrm{pipe}}=\min(\lambda_r,\lambda_t).
\end{equation}
This objective balances rollout production and trainer consumption
under a fixed resource budget, rather than maximizing either stage rate in isolation.
The selected partition reduces structural bubbles, while the compatible layouts make later bidirectional scheduling feasible.

\begin{algorithm}[t]
\small
\caption{Scheduling-aware static planning}
\label{alg:static-planning}
\begin{algorithmic}[1]
\Require device budget \(G\), placement constraint \(\Pi\), workload statistics \(W=(B_p,U_B,\ldots)\), model configuration \(P\)
% \Ensure resource partition, replica layouts, and stage-time models
% \State \(\rho_r \gets \Call{MinFeasibleLayout}{\textsc{Rollout},W,P}\)
% \State \(\rho_t \gets \Call{MinFeasibleLayout}{\textsc{Trainer},W,P}\)
\State \(\rho_r,\rho_t \gets \Call{MinFeasibleLayouts}{W,P}\)
\State \(\mathcal{M}_r,\mathcal{M}_t \gets \Call{BuildStageModels}{W,P,\rho_r,\rho_t}\)
\State \(best \gets \bot\)
\ForAll{\((g_r,g_t) \in \Call{CandidatePartitions}{G,\Pi}\)}
  \If{\(\neg\Call{Compatible}{g_r,g_t,\rho_r,\rho_t}\)}
    \State \textbf{continue}
  \EndIf
  \State \(d_r \gets \Call{ReplicaCount}{g_r,\rho_r}\)
  \State \(d_t \gets \Call{ReplicaCount}{g_t,\rho_t}\)
  \State \(T_r \gets \mathcal{M}_r(B_p,d_r)\)
  \State \(T_t \gets \mathcal{M}_t(U_B,d_t)\)
  \State \(\lambda_r \gets B_p/T_r\)
  \State \(\lambda_t \gets B_p/T_t\)
  \State \(\lambda_{\mathrm{pipe}} \gets \min(\lambda_r,\lambda_t)\)
  \If{\(best=\bot\) or \(\lambda_{\mathrm{pipe}} > best.\lambda_{\mathrm{pipe}}\)}
    \State \(best \gets (g_r,g_t,d_r,d_t,\lambda_{\mathrm{pipe}})\)
  \EndIf
\EndFor
\State \Return \(best.g_r,best.g_t,\rho_r,\rho_t,\mathcal{M}_r,\mathcal{M}_t\)
\end{algorithmic}
\end{algorithm}

Algorithm~\ref{alg:static-planning} summarizes the procedure,
where \(g\) is the number of devices in a pool and \(d\) is the number of replicas induced by the selected layout.
The planner returns the selected resource envelope,
which is then made executable by the hot-switch runtime and used by the bidirectional scheduler for admission control and workload splitting.

\subsection{Planning Complexity}
Let \(\mathcal{S}_{\Pi}\) denote the set of candidate rollout-training partitions generated under the placement constraint \(\Pi\).
Algorithm~\ref{alg:static-planning} scans each candidate partition once.
For each partition, the planner performs compatibility checks against the selected replica layouts,
computes the rollout and trainer replica counts, and evaluates the two stage-time models, each of which takes constant time.
Thus, after the one-time construction of replica layouts and stage models,
the partition-search complexity is \(O(|\mathcal{S}_{\Pi}|)\).
For the common case of two-way integer partitions over \(G\) devices,
\(|\mathcal{S}_{\Pi}|=O(G)\).
Since static planning runs once before RL training starts,
this overhead is amortized over the full training execution.
\section{Bidirectional Scheduling}
\label{sec:bidirectional-scheduling}

% BiDiRL reduces idle time through a planning-scheduling co-design.
% Before execution, the scheduling-aware static planner selects a throughput-aware and hot-switch-compatible resource envelope,
% including the initial rollout-training resource split and the replica layouts that define valid switching granularity.
% This planning step reduces \emph{structural bubbles} by balancing the average rates of rollout and training.
% During execution, the bidirectional scheduler uses this envelope and the exported stage-time models to reduce \emph{residual bubbles},
% left by staleness constraints, workload variation, and rate unmatching.
% BiDiRL preserves the disaggregated execution model: rollout and training still overlap on committed resource pools,
% while an idle resource pool can temporarily enter its auxiliary role when the predicted step-time reduction outweighs the hot-switch overhead.

% The coordinator enforces exclusive pool ownership. Trainer chunks that have not started execution can return to the pending queue.
% Running chunks finish and merge in sample order. Rollout groups either commit as complete groups or return as partial groups.
% These rules preserve the logical RL interface while allowing the runtime to reclaim idle windows (Appendix~\ref{app:chunk-pipeline}).

\subsection{Online Profiler}\label{sec:online-profiler}
% Online profiling checks the calibrates the runtime models used by the scheduler.
% The profiler does not only maintain aggregate throughput. It records per-replica execution summaries that capture three classes of information: the assigned workload,
% the observed execution time, and the ownership context.  The workload summary includes batch size and token-length statistics.
% The ownership context identifies whether the work ran on primary or auxiliary workers and which pool executed it.
The online profiler exposes a compact runtime interface to the bidirectional scheduler.
Each completed execution unit emits \((k,x,\pi,X,d,\hat{T})\):
\(k\in\{r,t\}\) is the rollout or trainer stage,
\(x\in\{p,a\}\) is the primary or auxiliary role,
\(\pi\) is the executing pool,
\(X\) summarizes the workload, \(d\) is the active replica count, and \(\hat{T}\) is the observed time.
Rollout records include prompt-group and token-length statistics, while trainer records include stage-ordered chunk workload.
These records calibrate \(\mathcal{M}_r\) and \(\mathcal{M}_t\).
The profiler also reports \((I_r,I_t,Q_r,U,C_{\mathrm{in}},C_{\mathrm{out}},C_{\mathrm{grad}})\),
where \(I_r,I_t\) are RollPoll and TrainPoll idle windows,
\(Q_r\) is the rollout deficit, \(U\) is the remaining trainer workload,
and the three \(C\) terms are measured switch-in, switch-out, and gradient-synchronization costs.
Workers emit records after completion, and calibration runs at weight-synchronization boundaries outside the worker critical path.

\subsection{Dynamic Scheduling}\label{sec:dynamic-scheduling}

The bidirectional scheduler reduces residual bubbles
by temporarily borrowing idle resources across the two committed resource pools.
Borrowing is considered when \(I_t>0\) for Rollouter-on-TrainPoll or \(I_r>0\) for Trainer-on-RollPoll.
Given the profiled hot-switch costs and stage-time models above, each borrowing opportunity makes two decisions:
whether the predicted benefit covers the hot-switch overhead, and how to split the remaining workload between primary and auxiliary workers if borrowing is admitted.
We use \(d_p\) and \(d_a\) to denote the numbers of primary and auxiliary replicas available to the bottleneck stage in the current borrowed window.

\subsubsection{Scheduling for Rollouter-on-TrainPoll.}
This direction uses idle training-pool resources as auxiliary rollouters to reduce trainer waiting time.
Let \(Q_r\) be the current deficit in prompt-level rollout groups required to start training under the staleness bound.
Let \(C_{\mathrm{in}}\) and \(C_{\mathrm{out}}\) be the measured costs for entering and leaving the auxiliary rollout window.
When the training pool is idle,
the scheduler admits borrowing only when primary rollouters alone are predicted to
need longer than the switching cost to close the deficit:
\begin{equation}
\label{eq:rot-gate}
\mathcal{M}_r(Q_r,d_p) - \mathcal{M}_r(Q_r,d_p+d_a)
>
C_{\mathrm{in}}+C_{\mathrm{out}}.
\end{equation}
Here \(\mathcal{M}_r\) is the rollout time model,
initialized by the static planner and calibrated by online profiling (Appendix~\ref{app:rollout-model}).
% The gate is conservative. If primary rollouters can close the deficit within the switching cost,
% the auxiliary window is not be opened.

After admission, BiDiRL dispatches the deficit according to rollout data-parallel capacity.
The auxiliary rollouters receive the following number of prompt groups:
\begin{equation}
\label{eq:rot-dispatch}
Q_a=
\operatorname{round}\left(\frac{d_a}{d_p+d_a}Q_r\right),
\qquad Q_p=Q_r-Q_a .
\end{equation}
The remaining \(Q_p\) groups stay on primary rollouters.
This design separates admission from per-request assignment.
The rollout time model \(\mathcal{M}_r\) predicts the aggregate production time of the deficit window,
while the final length of each individual response is only observed after generation.
BiDiRL therefore assigns prompts according to data-parallel capacity
rather than relying on fine-grained length prediction before generation.

Auxiliary rollout remains useful even when the borrowed window is interrupted.
A complete prompt group is committed to training immediately.
An incomplete group is returned to primary rollouters with its generated prefix and per-response completion status.
Primary rollouters resume unfinished responses rather than regenerating completed prefixes.
Similar to partial rollout, this design allows auxiliary rollout work to be preempted.

\subsubsection{Scheduling for Trainer-on-RollPoll}
Trainer-on-RollPoll reduces trainer consumption time by temporarily using RollPoll resources as auxiliary trainers.
Let \(U\) be the remaining stage-ordered trainer workload.
The trainer workload is decomposed into a sequence of chunk-level work.
A chunk is the basic execution unit submitted to trainer workers and is handled by primary trainers by default.
When RollPoll is idle, the scheduler admits borrowing
only when joint execution with auxiliary trainers
is predicted to finish the remaining chunks faster than primary-only execution.
Before admission, a startup drain removes the prefix of \(U\) that primary trainers are expected to finish while auxiliary trainers are being prepared,
yielding a drained workload \(U'\).
The admission test compares primary-only execution on \(U\) with joint execution on \(U'\) plus switching and gradient-synchronization costs.

After admission, the scheduler first assigns chunks in a demand-driven manner.
Primary and auxiliary trainers pull the next chunk only when they are ready to execute it.
This lazy-pull policy reduces sensitivity to variable chunk execution time
compared with one-time static assignment.
When the remaining chunk-level work fits within one combined primary and auxiliary window,
the scheduler performs a one-time tail split that searches over feasible chunk assignments
and uses the trainer time model (Appendix~\ref{app:trainer-model}) to choose the split that minimizes the predicted joint completion time.
% a one-time tail split is performed to fine-tune the final workload balance,
% where the chunk assignment is decided by the trainer model to minimize the joint completion time in a search way.
% The final optimizer step remains under primary trainers, and auxiliary trainers just contribute the gradients.

Chunk-level scheduling gives the runtime more opportunities to recover idle rollout resources,
but it introduces a trade-off between execution efficiency and scheduling granularity.
Smaller chunks improve tail alignment but reduce computation efficiency,
while larger chunks improve efficiency but may leave a larger final imbalance.
BiDiRL addresses this trade-off by using large chunks during lazy pull and allowing smaller remainder chunks only in the tail partitioning.
Data-transfer overheads from chunk-level execution are further hidden by the asynchronous input-compute-output pipeline described in Section~\ref{para:chunk-pipeline}.
% There is also a trade-off in chunk size choice. Too small chunks decrease the computation efficiency,
% while too large chunks may result in tail misalignment between primary and auxiliary trainers.
% BiDiRL balances this trade-off with lazy pull using as large chunks as possible for efficiency,
% and tail split using more small chunks if needed.

Algorithm~\ref{alg:trainer-on-rollout} summarizes Trainer-on-RollPoll scheduling. Here \(\mathcal{Q}\) tracks the remaining chunk-level work rather than a fixed list of concrete chunks.
\(\Call{LazyPull}{\mathcal{Q},x,s_{\max}}\) forms one ready executable chunk of size up to \(s_{\max}\) for side \(x\).
\(\Call{JointTailSplit}{}\) re-chunks the remaining tail work for the primary and auxiliary sides separately.
For each candidate workload split, both sides use full chunks of size \(s_{\max}\) and at most one smaller remainder chunk.
The minimum chunk size is one sample per device.
The selected workload split minimizes the larger predicted completion time of the two sides.
\(C_{\mathrm{grad}}\) is nonzero only when the borrowed window covers actor update and auxiliary gradients must be merged.
% It omits the Ray event loop and worker-side buffers,
% which are implementation details described in Section~\ref{sec:implementation} and Appendix~\ref{app:chunk-pipeline}.

\begin{algorithm}[t]
\small
\caption{Trainer-on-RollPoll scheduling}
\label{alg:trainer-on-rollout}
\begin{algorithmic}[1]
\Require stage-ordered workload \(U\), replica counts \(d_p,d_a\), maximum chunk size \(s_{\max}\), costs \(C_{\mathrm{in}},C_{\mathrm{out}},C_{\mathrm{grad}}\)
\State \(U'\gets\Call{DrainPrefix}{U,C_{\mathrm{in}},\mathcal{M}_t,d_p}\)
\State \(T_{\mathrm{no}}\gets\mathcal{M}_t(U,d_p)\)
\State \(T_{\mathrm{joint}}\gets\mathcal{M}_t(U',d_p+d_a)\)
\State \(T_{\mathrm{borrow}}\gets C_{\mathrm{in}}+T_{\mathrm{joint}}+C_{\mathrm{out}}+C_{\mathrm{grad}}\)
\If{\(T_{\mathrm{borrow}}\ge T_{\mathrm{no}}\)}
  \State \Return primary-only execution
\EndIf
% \State \Call{PrepareAuxTrainer}{}
\State \(\mathcal{Q}\gets\Call{BuildChunkQueue}{U'}\)
\While{\(\mathcal{Q}\) is not empty}
  \If{\(\Call{TailWindow}{\mathcal{Q}}\)}
    \State \((U_p,U_a)\gets\Call{JointTailSplit}{\mathcal{Q},s_{\max},\mathcal{M}_t,d_p,d_a}\)
    \State \(\Call{RunTail}{U_p,U_a}\)
    \State \textbf{break}
  \EndIf
  \ForAll{ready side \(x\in\{p,a\}\)}
    \State \(c\gets\Call{LazyPull}{\mathcal{Q},x,s_{\max}}\)
    \State \Call{RunAsync}{x,c}
  \EndFor
\EndWhile
% \If{auxiliary update gradients exist}
%   \State \Call{MergeGradients}{}
% \EndIf
% \State \Call{ReleaseAuxTrainer}{}
\end{algorithmic}
\end{algorithm}
% The final optimizer step remains under primary trainers, and auxiliary trainers just contribute the gradients.

% \textsc{DrainPrefix} removes only chunks that primary trainers are predicted to finish during auxiliary preparation.
% This avoids charging the borrowed window for work that would have completed before auxiliary trainers become available.
% The admission test then compares primary-only execution with joint execution on the drained workload.
% \(C_{\mathrm{grad}}\) is nonzero only when the borrowed window covers actor update and auxiliary gradients must be merged.

% During the main phase, lazy pull delays ownership until a primary or auxiliary trainer is ready to execute the next chunk.
% This avoids committing the entire remaining workload based on an early prediction. When the remaining work fits in one combined primary/auxiliary window,
% the scheduler performs a joint tail split.  It chooses the split that minimizes the predicted joint completion time using the calibrated trainer models.
% The tail split is performed once.

\subsubsection{Scheduling Complexity}
The Rollouter-on-TrainPoll decision incurs constant scheduler-side overhead per borrowing opportunity,
as it only evaluates one rollout-time prediction and computes a capacity-proportional dispatch.
For Trainer-on-RollPoll, let \(n\) be the number of chunks actually dispatched by lazy pull.
Each dispatched chunk is pulled and launched once, so the lazy-pull loop has \(O(n)\) queue-management overhead.
The tail split only enumerates integer split points within the bounded tail window.
If the tail window contains \(u\) minimum scheduling units, the tail-split search costs \(O(u)\).
% The scheduler only constructs chunk descriptors during this search and does not execute candidate chunks.
Thus, the scheduler-side overhead is negligible compared with the model execution time.

\section{Implementation}
\label{sec:implementation}

\begin{table}[t]
\centering
% \scriptsize
\small
\caption{Trainer chunk state classes used for recovery and ordered merge.}
\begin{tabular}{@{}>{\raggedright\arraybackslash}p{0.22\columnwidth}
                >{\raggedright\arraybackslash}p{0.44\columnwidth}
                >{\raggedright\arraybackslash}p{0.22\columnwidth}@{}}
\toprule
State & Meaning & Recoverable \\
\midrule
\texttt{queued} & Input transfer or worker-side buffering has started, but accelerator execution has not started. & Yes \\
\texttt{running} & The worker is executing the chunk on the accelerator. & No \\
\texttt{finished} & Execution output has been retrieved and merged in original order. & No \\
\texttt{cancelled} & A queued chunk was released before execution. & Terminal \\
\bottomrule
\end{tabular}
\label{tab:chunk-states}
\end{table}

\subsection{Prototype}
We implemented \sys{} on top of veRL~\cite{shengHybridFlowFlexibleEfficient2025} in 8.3K lines of Python code.
\sys{} uses vLLM~\cite{kwonEfficientMemoryManagement2023} for rollout, PyTorch FSDP~\cite{zhaoPyTorchFSDPExperiences2023}
for training, and Ray~\cite{moritzRayDistributedFramework} for orchestration. 
% BiDiRL keeps the algorithm-facing dataflow of verl unchanged. 
% Rollout groups, training batches, and actor-update calls are exposed to the RL algorithm in the same form as in the original stack. 
We develop the logic for static resource planning, bidirectional scheduling, workload partitioning and execution, worker lifecycle management,
hot switching, cross-pool synchronization, asynchronous chunk execution,
and online profiling around the existing rollout and training execution paths.

\subsection{Cross-pool Synchronization}
\sys{} reuses the veRL-style actor-to-rollout synchronization path between primary trainers and primary rollouters.
It adds three synchronization paths for borrowed execution. 
In Trainer-on-RollPoll, the canonical primary trainer broadcasts the required actor weights 
to all auxiliary trainers on the rollout pool to refresh their model weights. 
In Rollouter-on-TrainPoll, the canonical primary rollouter also broadcasts rollout weights 
to auxiliary rollouters on the training pool. 
If auxiliary trainers participate in actor update, their gradients are synchronized back to the primary trainer before the optimizer step.
These synchronization paths rely on the replica compatibility selected by the static planner (Appendix~\ref{app:planner-models}).
Primary and auxiliary workers of the same stage use compatible replica configurations even when they run on different pools,
so hot switching does not require process restart or runtime resharding~\cite{leiPuzzleEfficientlyAligning}.

\subsection{Asynchronous Trainer Chunks}
\label{para:chunk-pipeline}
Trainer-on-RollPoll executes training work through an asynchronous chunk pipeline; Table~\ref{tab:chunk-states} summarizes the recoverable chunk states.
Each chunk follows an enqueue-run-dequeue sequence.
The enqueue stage transfers the chunk input to the target worker and places it in a worker-side buffer.
The run stage launches the corresponding trainer operation on the accelerator.
The dequeue stage retrieves the completed output and merges it according to the original chunk order.
Dependency-free operations, including Ray object transfers, worker-side buffering, chunk execution,
output retrieval, and auxiliary preparation or release events, are issued asynchronously.
This allows transfers, execution, and lease-management events from different chunks to overlap.
The scheduler may therefore keep multiple chunk futures in flight across primary and auxiliary trainers.

The pipeline hides the overhead of chunk-level scheduling while preserving trainer-side ordering semantics.
Chunks that have only been enqueued or buffered can be reclaimed when the auxiliary lease is revoked.
Chunks that are already running are allowed to finish to avoid wasting accelerator resources.
Completed chunks are dequeued and merged in the original order, making the chunk pipeline behave as if
the chunks were executed sequentially on the primary trainer.

\section{Evaluation}
\label{sec:evaluation}

\begin{figure*}[t]
  \centering
  \captionsetup[subfigure]{font=small,justification=centering,singlelinecheck=false}
  \includegraphics[width=0.42\textwidth]{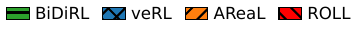}
  \vspace{0.1em}

  \begin{subfigure}[t]{0.24\textwidth}
    \centering
    \includegraphics[width=\linewidth]{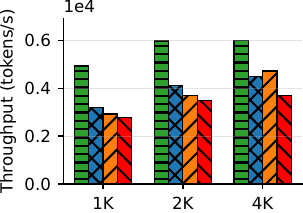}
    \caption{Response length\protect\\[-0.2ex]{\scriptsize(\(1.27\times\)--\(1.55\times\))}}
    \label{fig:eval-main-a}
  \end{subfigure}
  \hfill
  \begin{subfigure}[t]{0.24\textwidth}
    \centering
    \includegraphics[width=\linewidth]{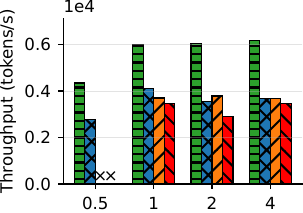}
    \caption{Staleness\protect\\[-0.2ex]{\scriptsize(\(1.45\times\)--\(1.68\times\))}}
    \label{fig:eval-main-b}
  \end{subfigure}
  \hfill
  \begin{subfigure}[t]{0.24\textwidth}
    \centering
    \includegraphics[width=\linewidth]{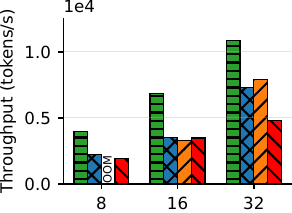}
    \caption{Resource (batch = 128)\protect\\[-0.2ex]{\scriptsize(\(1.38\times\)--\(1.94\times\))}}
    \label{fig:eval-main-c}
  \end{subfigure}
  \hfill
  \begin{subfigure}[t]{0.24\textwidth}
    \centering
    \includegraphics[width=\linewidth]{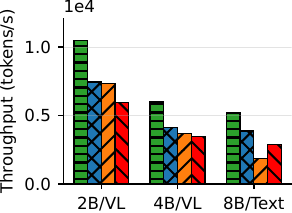}
    \caption{Model/dataset\protect\\[-0.2ex]{\scriptsize(\(1.34\times\)--\(1.45\times\))}}
    \label{fig:eval-main-d}
  \end{subfigure}

  \vspace{0.25em}

  \begin{subfigure}[t]{0.24\textwidth}
    \centering
    \includegraphics[width=\linewidth]{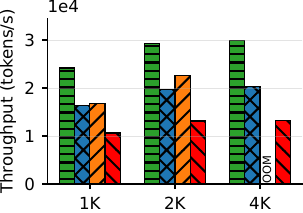}
    \caption{\(\ddagger\) Response length\protect\\[-0.2ex]{\scriptsize(\(1.29\times\)--\(1.47\times\))}}
    \label{fig:eval-main-h100-a}
  \end{subfigure}
  \hfill
  \begin{subfigure}[t]{0.24\textwidth}
    \centering
    \includegraphics[width=\linewidth]{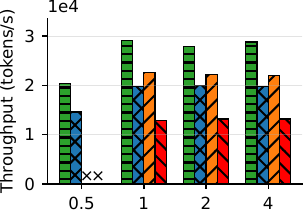}
    \caption{\(\ddagger\) Staleness\protect\\[-0.2ex]{\scriptsize(\(1.26\times\)--\(1.39\times\))}}
    \label{fig:eval-main-h100-b}
  \end{subfigure}
  \hfill
  \begin{subfigure}[t]{0.24\textwidth}
    \centering
    \includegraphics[width=\linewidth]{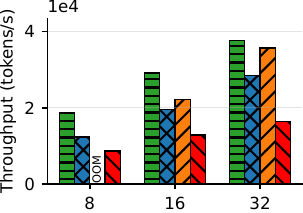}
    \caption{\(\ddagger\) Resource (batch = 128)\protect\\[-0.2ex]{\scriptsize(\(1.05\times\)--\(1.53\times\))}}
    \label{fig:eval-main-h100-c}
  \end{subfigure}
  \hfill
  \begin{subfigure}[t]{0.24\textwidth}
    \centering
    \includegraphics[width=\linewidth]{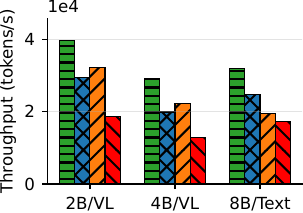}
    \caption{\(\ddagger\) Model/dataset\protect\\[-0.2ex]{\scriptsize(\(1.23\times\)--\(1.31\times\))}}
    \label{fig:eval-main-h100-d}
  \end{subfigure}

  \vspace{0.25em}

  \begin{subfigure}[t]{0.24\textwidth}
    \centering
    \includegraphics[width=\linewidth]{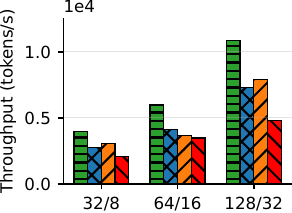}
    \caption{Batch/resource scaling\protect\\[-0.2ex]{\scriptsize(\(1.29\times\)--\(1.45\times\))}}
    \label{fig:eval-main-resource-a}
  \end{subfigure}
  \hfill
  \begin{subfigure}[t]{0.24\textwidth}
    \centering
    \includegraphics[width=\linewidth]{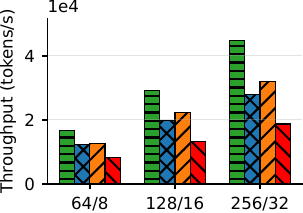}
    \caption{\(\ddagger\) Batch/resource scaling\protect\\[-0.2ex]{\scriptsize(\(1.30\times\)--\(1.41\times\))}}
    \label{fig:eval-main-resource-b}
  \end{subfigure}
  \hfill
  \begin{subfigure}[t]{0.24\textwidth}
    \centering
    \includegraphics[width=\linewidth]{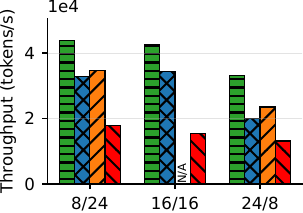}
    \caption{\(\ddagger\) Resource partition (4B)\protect\\[-0.2ex]{\scriptsize(\(1.24\times\)--\(1.41\times\))}}
    \label{fig:eval-main-resource-c}
  \end{subfigure}
  \hfill
  \begin{subfigure}[t]{0.24\textwidth}
    \centering
    \includegraphics[width=\linewidth]{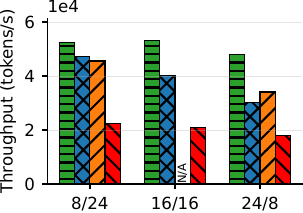}
    \caption{\(\ddagger\) Resource partition (2B)\protect\\[-0.2ex]{\scriptsize(\(1.11\times\)--\(1.41\times\))}}
    \label{fig:eval-main-resource-d}
  \end{subfigure}

  \caption{
  End-to-end throughput comparison across A6000/H100 workloads and resource settings.
  Bars show raw throughput; speedup ranges compare BiDiRL with the strongest valid external baseline.
  OOM, failed, and unsupported runs are excluded; \(\times\) marks unsupported settings and \(\ddagger\) marks H100 results.
  }
  \label{fig:eval-main}
\end{figure*}

BiDiRL improves disaggregated LLM RL throughput by harvesting residual idle windows between rollout and training through model-guided bidirectional scheduling.
The evaluation shows that this benefit holds across workload, staleness, resource, model, dataset, and hardware changes, while preserving the logical RL dataflow.
We first compare end-to-end throughput against state-of-the-art LLM RL systems.
We then isolate the scheduling mechanisms, validate the static planning and stage-time models that guide scheduling decisions, and check training behavior.

\subsection{Experimental Setup}
\label{sec:eval-setup}

\begin{table}[!t]
  \centering
  \caption{Per-server configuration of the A6000 testbed.}
  \label{tab:a6000-server-config}
  \footnotesize
  \begin{tabular}{@{}p{0.18\columnwidth}p{0.74\columnwidth}@{}}
    \toprule
    Component & Configuration \\
    \midrule
    CPU & Dual Intel(R) Xeon(R) Platinum 8358 @ 2.60GHz \\
    GPU & 8x NVIDIA RTX A6000-48G @ 1.46GHz \\
    Memory & 512GB DDR4 \\
    NVLink & 112.5GB/s (4x) \\
    PCIe & 4.0 (x16) \\
    Network & Mellanox MT28908 @ 200Gb/s \\
    \bottomrule
  \end{tabular}
\end{table}

\begin{table}[!t]
  \centering
  \caption{Per-server configuration of the H100 testbed.}
  \label{tab:h100-server-config}
  \footnotesize
  \begin{tabular}{@{}p{0.18\columnwidth}p{0.74\columnwidth}@{}}
    \toprule
    Component & Configuration \\
    \midrule
    CPU & Dual Intel(R) Xeon(R) Platinum 8468 @ 2.10GHz \\
    GPU & 8x NVIDIA H100-80G HBM3 @ 1.98GHz \\
    Memory & 2TB DDR5 \\
    NVLink & \(\sim\)956GB/s (18x) \\
    PCIe & 5.0 (x16) \\
    Network & 8x Mellanox MT2910 @ 400Gb/s \\
    \bottomrule
  \end{tabular}
\end{table}

\textbf{Testbed. }
Unless otherwise specified, experiments are run on the A6000 GPU cluster which contains four nodes (the node configuration is shown in Table~\ref{tab:a6000-server-config}).
The H100 testbed also contains a total of 32 GPUs with four nodes (Table~\ref{tab:h100-server-config}), which are used for cross-hardware validation and are explicitly marked in the figures.
We evaluate BiDiRL across scenarios with max response lengths from 1K to 4K, staleness bounds from 0 to 4, 
both text and multimodal datasets, model sizes from 2B to 8B, and resource budgets from 8 to 32 devices.
Specifically, we use models from the Qwen 3 family~\cite{yangQwen3TechnicalReport2025} including Qwen3VL-2B, Qwen3VL-4B, and Qwen3-8B. 
The multimodal dataset is Geo3K~\cite{lu-etal-2021-inter} and the text dataset is GSM8K~\cite{cobbe2021gsm8k}. 
Unless otherwise specified, we use Geo3K with a max response length of 2K and a global prompt batch size of 64 as the default workload, 
with a GRPO group size of 8 responses per prompt. The batch size is doubled on H100 devices to better use the larger memory capacity. 
By default, the staleness bound is 1, the model is Qwen3VL-4B, and the resource budget is 2 nodes.

\textbf{Baselines. }
We compare BiDiRL with veRL v0.7.1~\cite{shengHybridFlowFlexibleEfficient2025}, AReaL v1.0.3~\cite{fuAReaLLargeScaleAsynchronous2025}, and ROLL v0.2.1~\cite{wangReinforcementLearningOptimization2025}.
veRL~\cite{shengHybridFlowFlexibleEfficient2025} provides practical colocated and disaggregated data flows for LLM RL.
AReaL~\cite{fuAReaLLargeScaleAsynchronous2025} pushes disaggregation toward fully asynchronous execution, 
where rollouters continuously generate samples and trainers update the actor once enough samples are available.
ROLL~\cite{wangReinforcementLearningOptimization2025} provides a large-scale RL system stack with rollout scheduling, parallel strategy management, and data-transfer support. 
We configure all systems with the same hyperparameter settings whenever supported.
Unsupported settings are marked as N/A.  Runs that fail due to memory pressure are marked as OOM and are excluded from speedup ranges. 
Fractional staleness values between 0 and 1 are unsupported by ROLL and AReaL, so those points are marked with \(\times\) in the staleness sweep.
AReaL also requires the batch size to be divisible by rollout data parallelism.
In the 2B and 4B H100 resource-partition sweeps, the batch-120 16-16 point violates this rule and is marked unsupported.

\textbf{Metrics. }
We compute RL training throughput as the total number of prompt and response tokens consumed by the measured training steps, divided by their step time. 
% For end-to-end comparisons, 
An experiment with staleness bound \(s\) drops \(\max(1,\lceil s\rceil)\) complete metric steps from the head and \(\lceil s\rceil\) steps from the tail 
for stability. 
% Throughput is the token sum of the selected steps divided by their time; the number of selected steps is 6 in our experiments.
% For ablation and static-split sweeps, all runs are BiDiRL variants under the same parser, so we instead report total-run throughput over all complete metric steps.
% The text reports normalized speedups over the strongest valid comparator rather than absolute throughput.

\subsection{End-to-End Performance}
\label{sec:eval-e2e}
Figure~\ref{fig:eval-main} summarizes the end-to-end throughput comparison.
For compatibility with the baselines, BiDiRL uses the same node-aligned rollout/trainer partitioning as the compared systems in the main end-to-end runs.
The static planner supplies the hot-switch-compatible envelope and stage-time models used by the runtime scheduler.

\begin{figure*}[t]
  \centering
  \captionsetup[subfigure]{font=small,justification=centering,singlelinecheck=false}
  \includegraphics[width=0.62\textwidth]{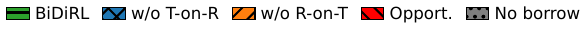}
  \vspace{0.2em}

  \begin{subfigure}[t]{0.24\textwidth}
    \centering
    \includegraphics[width=\linewidth]{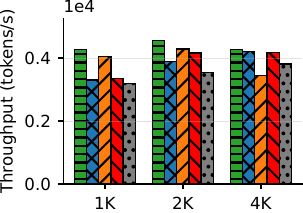}
    \caption{Response length\protect\\[-0.2ex]{\scriptsize(\(1.02\times\)--\(1.06\times\))}}
    \label{fig:eval-ablation-a}
  \end{subfigure}
  \hfill
  \begin{subfigure}[t]{0.24\textwidth}
    \centering
    \includegraphics[width=\linewidth]{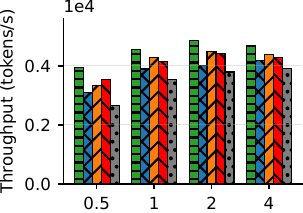}
    \caption{Staleness\protect\\[-0.2ex]{\scriptsize(\(1.06\times\)--\(1.12\times\))}}
    \label{fig:eval-ablation-b}
  \end{subfigure}
  \hfill
  \begin{subfigure}[t]{0.24\textwidth}
    \centering
    \includegraphics[width=\linewidth]{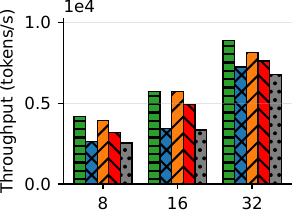}
    \caption{Resource (batch = 120)\protect\\[-0.2ex]{\scriptsize(\(1.00\times\)--\(1.09\times\))}}
    \label{fig:eval-ablation-c}
  \end{subfigure}
  \hfill
  \begin{subfigure}[t]{0.24\textwidth}
    \centering
    \includegraphics[width=\linewidth]{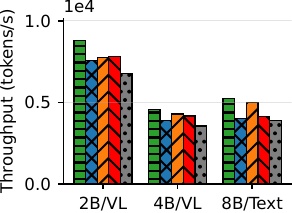}
    \caption{Model/dataset\protect\\[-0.2ex]{\scriptsize(\(1.04\times\)--\(1.13\times\))}}
    \label{fig:eval-ablation-d}
  \end{subfigure}

  \caption{
  A6000 ablation throughput summary. The figure isolates the contribution of model-guided bidirectional scheduling.
  Bars show raw throughput; speedup ranges compare BiDiRL with the strongest valid ablated variant, excluding failed runs.
  }
  \label{fig:eval-ablation}
\end{figure*}

\textbf{Overall performance. }
As shown in Figure~\ref{fig:eval-main}, BiDiRL runs faster than existing veRL, AReaL, and ROLL by $1.05\times$--$1.94\times$ under different configurations.
On A6000, it improves over the baselines by \(1.27\times\)--\(1.68\times\) across the default, staleness, batch/resource, and model/dataset sweeps.
The scale-up setting reaches \(1.94\times\).
On H100, BiDiRL increases throughput by \(1.23\times\)--\(1.47\times\) in the default setting and \(1.05\times\)--\(1.53\times\) in scale-up.
The H100 resource-partition studies give \(1.11\times\)--\(1.41\times\) speedups.
Existing systems keep fixed committed rollout and trainer pools within an execution window, so asynchronous overlap still leaves residual idle windows when the two rates diverge.
BiDiRL turns these windows into useful work by temporarily lending idle resources to the current bottleneck stage.

\textbf{Workload and staleness. }
BiDiRL does not rely on a fixed bottleneck direction.
Response length, model size, and dataset shift the relative cost of rollout production and trainer consumption.
On A6000, BiDiRL is \(1.55\times\), \(1.45\times\), and \(1.27\times\) faster for 1K, 2K, and 4K responses, respectively.
It is also \(1.34\times\)--\(1.45\times\) faster in the model/dataset sweep.
The smaller gains occur when sample production and consumption are already closer to balanced, or when memory-safe configurations reduce reusable slack.
The staleness sweep stresses the other part of the design.
Tight bounds make trainers wait for enough new rollout groups, while relaxed bounds allow more natural overlap but still leave bubbles when production and consumption rates mismatch.
At \(s=0\), we separately compare BiDiRL with the veRL colocated architecture.
With bidirectional borrowing, pipeline bubbles~\cite{zhongStreamRLScalableHeterogeneous} are largely eliminated, so the two systems achieve similar throughput.
For \(s>0\), BiDiRL improves staleness throughput by \(1.45\times\)--\(1.68\times\) on A6000 and \(1.26\times\)--\(1.39\times\) on H100.
This shows that bidirectional scheduling remains useful whether the exposed idle window appears on the trainer side, the rollout side, or both.

\textbf{Resource and hardware. }
With different resource scaling up, the runtime dynamics become more severe and the runtime scheduling is more important.
On A6000, BiDiRL improves the 128-prompt setting by \(1.38\times\)--\(1.94\times\) across 8 to 32 devices.
On H100, the same pattern illustrates that our \sys{} achieves \(1.05\times\)--\(1.53\times\) speedups, and the resource-partition studies give \(1.24\times\)--\(1.41\times\) for Qwen3VL-4B and \(1.11\times\)--\(1.41\times\) for Qwen3VL-2B.
The batch/resource setting further shows that changing batch size together with device count does not make a fixed partition universally balanced.
Larger or faster hardware shifts the rollout/trainer rate ratio, but residual imbalance remains under fixed disaggregated pools. 
The relative speedup is smaller on H100 because the stronger devices and larger batches leave shorter residual idle windows for BiDiRL to harvest.
BiDiRL adapts by borrowing in the direction whose predicted gain exceeds switching overhead.

\subsection{Scheduling Ablation}
\label{sec:eval-ablation}

Figure~\ref{fig:eval-ablation} evaluates the scheduling mechanisms.
The end-to-end comparison uses the default 8-8 partition because it is a common disaggregated configuration across the compared systems.
For ablation, we use a 4-12 rollout/trainer partition under the same 16-device budget and prompt batch size 60 for the multimodal workload.
% This partition intentionally stresses the scheduler.
As response length and staleness change, it exposes both rollout-heavy and trainer-heavy windows, making it suitable for testing both borrowing directions.
The Qwen3-8B/GSM8K ablation uses the 8-8 partition and prompt batch size 256 used by the corresponding end-to-end text run to avoid OOM.

\textbf{Ablation variants.}   
\emph{No borrow} keeps the same static partition but disables runtime borrowing.
\emph{w/o R-on-T} disables trainer-pool resources from being borrowed for rollout.
\emph{w/o T-on-R} disables rollout-pool resources from being borrowed for training.
\emph{Opport.} borrows whenever the opposite pool is observed idle, without checking whether the window covers switching overhead and without model-guided workload splitting.
BiDiRL consistently matches or exceeds the strongest valid ablated variant, with \(1.00\times\)--\(1.13\times\) speedup under the total-run throughput metric.
The gap against simpler policies is larger.
BiDiRL improves over no borrowing by \(1.12\times\)--\(1.71\times\), and over opportunistic borrowing by \(1.02\times\)--\(1.31\times\).
This pattern shows that the benefit comes from model-guided admission and bidirectional workload splitting, not only from enabling borrowing.
One-direction and opportunistic variants can be strong in particular states, but they become suboptimal when the workload or staleness state changes.

\textbf{Bidirectional scheduling.}
Both borrowing directions are necessary because different runtime states expose different idle pools.
Across the ablation sweeps, BiDiRL is \(1.02\times\)--\(1.68\times\) faster than \emph{w/o T-on-R} and \(1.00\times\)--\(1.24\times\) faster than \emph{w/o R-on-T}.
It is \(1.00\times\)--\(1.19\times\) faster than the stronger one-direction variant in each setting.
Longer responses make rollout heavier, so Rollouter-on-TrainPoll becomes important when trainers are waiting for enough valid rollout groups.
Shorter responses or relaxed staleness can instead leave rollout resources idle while trainers consume the batch, making Trainer-on-RollPoll more useful.
For example, \(s=0\) and \(0<s<1\) settings with faster rollout production can expose both \(T_{\mathrm{wait}}\) and \(T_{\mathrm{consume}}\) windows in the same run, which is why the full BiDiRL policy is more robust than either one-direction variant.
The ablation therefore supports the core design choice in Section~\ref{sec:bidirectional-scheduling}.
Residual bubbles are two-sided and cannot be handled by optimizing only rollout or only training.

\textbf{Model-guided admission and partitioning.}
Not every idle window is worth borrowing.
BiDiRL is \(1.02\times\)--\(1.31\times\) faster than opportunistic borrowing and \(1.12\times\)--\(1.71\times\) faster than no borrowing.
Opportunistic borrowing improves over no borrowing, but it still falls below BiDiRL.
Short windows may not cover hot-switch cost, and one-shot DP-proportional assignment can leave primary/auxiliary tail mismatch.
BiDiRL uses the stage-time model to admit a borrowed window only when the predicted reduction exceeds switching cost, and it splits the remaining work according to the predicted completion time of primary and auxiliary workers.
The result is a scheduler that reclaims slack when it is useful and avoids paying switching overhead for windows that are too short.

\begin{table}[t]
  \centering
  \caption{Measured hot-switch costs (in seconds).}
  \label{tab:hot-switch-cost}
  % \scriptsize
  % \setlength{\tabcolsep}{2.5pt}
  % \renewcommand{\arraystretch}{0.9}
  \begin{tabular}{@{}lccc@{}}
    \toprule
    Borrowing path & \(C_{\mathrm{in}}\) & \(C_{\mathrm{out}}\) & \(C_{\mathrm{grad}}\) \\
    \midrule
    Trainer-on-RollPoll & 3.58/6.16 & 3.40/5.62 & 0.66/1.21 \\
    Rollouter-on-TrainPoll & 5.54/7.70 & 4.20/5.59 & -- \\
    \bottomrule
  \end{tabular}
\end{table}

\textbf{Hot-switch cost.}
Table~\ref{tab:hot-switch-cost} reports the hot-switch cost components charged by the admission tests in Section~\ref{sec:bidirectional-scheduling}, with each entry shown as Qwen3VL-2B/Qwen3VL-4B seconds.
\(C_{\mathrm{in}}\) and \(C_{\mathrm{out}}\) cover model state movement and synchronization when a worker enters and leaves an auxiliary role, while \(C_{\mathrm{grad}}\) is charged only when Trainer-on-RollPoll performs actor-update synchronization.
The 4B costs are consistently higher than the 2B costs, as hot switching is dominated by model state size and cluster communication bandwidth.
These costs are small relative to rollout/training windows but non-negligible for short idle windows.
BiDiRL therefore admits borrowing only when the model-estimated gain exceeds the measured switching cost, rather than borrowing on every observed idle window.

\subsection{Static Planning and Model Validation}
\label{sec:eval-static}

\begin{figure}[t]
  \centering
  \begin{subfigure}[t]{0.47\columnwidth}
    \centering
    \includegraphics[width=\linewidth]{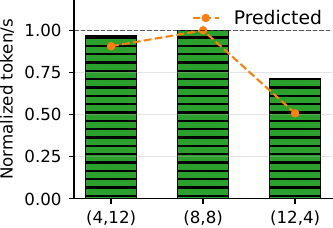}
    \caption{2B resource partition}
    \label{fig:static-a}
  \end{subfigure}
  \hfill
  \begin{subfigure}[t]{0.47\columnwidth}
    \centering
    \includegraphics[width=\linewidth]{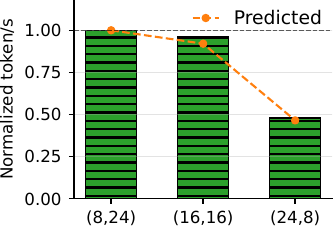}
    \caption{4B resource partition}
    \label{fig:static-b}
  \end{subfigure}

  \vspace{0.35em}

  \begin{subfigure}[t]{0.47\columnwidth}
    \centering
    \includegraphics[width=\linewidth]{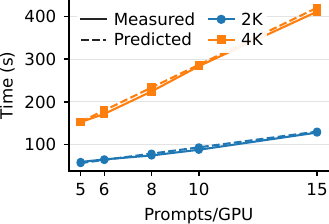}
    \caption{Rollout}
    \label{fig:static-c}
  \end{subfigure}
  \hfill
  \begin{subfigure}[t]{0.47\columnwidth}
    \centering
    \includegraphics[width=\linewidth]{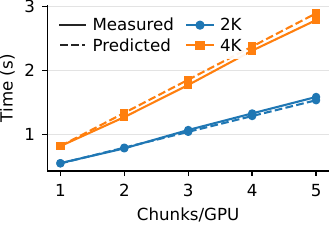}
    \caption{Trainer}
    \label{fig:static-d}
  \end{subfigure}

  \caption{
  Static planning and model validation.
  Partition sweeps compare measured throughput with planner predictions, where \((R,T)\) denotes rollout/trainer devices.
  The rollout and trainer panels compare measured and predicted stage time.
  }
  \label{fig:eval-static}
\end{figure}

\textbf{Static planning.}
Figure~\ref{fig:eval-static} validates the resource envelope selected by the static planner.
The partition sweeps show that static partitioning matters because different rollout/trainer partitions produce clearly different measured throughput under the same total resource budget.
Across the displayed machine-aligned partitions, measured throughput varies by \(1.40\times\) for Qwen3VL-2B and \(2.09\times\) for Qwen3VL-4B.
The predicted trend follows the measured trend and selects the measured-best partition in the displayed sweeps.
The selected partitions are 8-8 for Qwen3VL-2B and 8-24 for Qwen3VL-4B.
The planner is not intended to predict every point with exact absolute accuracy.
Its role is to avoid obviously imbalanced partitions and select a high-throughput, hot-switch-compatible resource envelope.
The best partition differs across the 2B and 4B models, confirming that model size changes the average stage-cost ratio.
This result complements the dynamic workload effects in Figure~\ref{fig:eval-main}.
Response length, dataset, and staleness further move the rollout/trainer rate ratio, so a fixed partition tuned for one condition is not sufficient across runs.

\textbf{Stage-time model validation.}
The stage-time models provide the ranking signal used by both static planning and runtime admission; Appendix~\ref{app:planner-models} gives their definitions.
Figure~\ref{fig:static-c} shows the rollout model and Figure~\ref{fig:static-d} shows the trainer model.
The rollout panel increases with prompts per GPU, the trainer panel increases with chunks per GPU, and the 2K and 4K response-length curves remain separated as expected.
The two fits have median errors of \(3.12\%\) and \(2.92\%\), with p90 errors of \(5.77\%\) and \(4.68\%\).
These errors are sufficient for ranking resource envelopes and deciding whether a borrowed window is likely to pay for its switching cost.

\subsection{Training Convergence}
\label{sec:eval-training}

\begin{figure}[t]
  \centering
  \includegraphics[width=0.86\columnwidth]{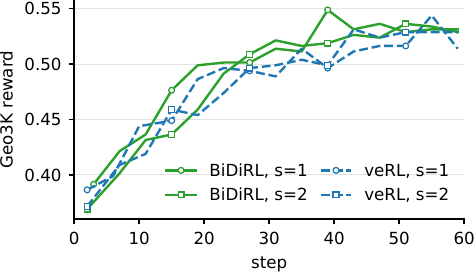}
  \caption{
  Convergence behavior over the first 60 training steps.
  BiDiRL changes computation placement and timing while preserving the logical rollout groups and training samples consumed by GRPO.
  }
  \label{fig:eval-convergence}
\end{figure}

Figure~\ref{fig:eval-convergence} compares BiDiRL with veRL under the same staleness settings.
BiDiRL changes the physical placement and timing of computation, but it preserves the logical rollout groups and training samples consumed by GRPO.
Within the first 60 steps, the last common logged-point reward gap is \(+0.017\) at staleness 1 and \(+0.000\) at staleness 2; the corresponding mean gaps are \(+0.020\) and \(+0.001\).
These small gaps show that bidirectional resource scheduling, as a system efficiency optimization, changes computation placement but does not alter the logical RL behavior exposed to GRPO.

\section{Discussion}
\label{sec:discussion}

\textbf{Static planning and runtime adaptation.}
Static planning is necessary but insufficient for LLM RL.
The workload distribution, especially response length, varies across prompts and evolves as the policy changes during training~\cite{deepseek-aiDeepSeekR1IncentivizingReasoning2025,liuDAPOImprovingMultiStep}.
BiDiRL therefore uses lightweight offline profiling only to select a hot-switch-compatible resource envelope and initialize the stage-time models, while online profiling corrects the models as the workload evolves.
Runtime speedups depend on the actual workload distribution and observed idle windows \(I_r\) and \(I_t\), which are only available during execution; BiDiRL therefore avoids offline prediction of borrowing gains.
Instead of repartitioning by stopping and restarting a training job, the online profiler continuously updates \(\mathcal{M}_r\), \(\mathcal{M}_t\), and switching costs within the selected envelope.

\textbf{Rollout sample granularity.}
Rollouter-on-TrainPoll schedules prompt groups.
BiDiRL splits the rollout deficit \(Q_r\) by replica capacity rather than predicting final response length, which is unknown before generation finishes.
This avoids relying on per-sample length prediction and keeps admission and dispatch simple, but an interrupted auxiliary window may return partially generated groups to primary rollouters.
Future schedulers could use prompt features or early decoding signals~\cite{heHistoryRhymesAccelerating2025} to place likely short prompts on auxiliary rollouters.

\textbf{Trainer chunk granularity.}
Trainer-on-RollPoll trades chunk efficiency for tail balance.
Large chunks improve accelerator efficiency, while small chunks reduce mismatch between primary and auxiliary trainers.
BiDiRL uses large lazy-pull chunks on the common path, then searches over the remaining workload split near the tail and uses smaller chunks only when needed to reduce primary/auxiliary mismatch.
Because auxiliary trainers compute gradients without optimizer states, future implementations can use the saved memory for more aggressive adaptive chunk sizing.

\section{Related Work}
\label{sec:related-work}

\textbf{LLM inference and training engines.}
LLM rollout and training have different execution characteristics, making it difficult for one engine to efficiently support both RL stages~\cite{zhongStreamRLScalableHeterogeneous,fuAReaLLargeScaleAsynchronous2025}.
Rollout is dominated by variable-length autoregressive generation and benefits from serving engines such as vLLM~\cite{kwonEfficientMemoryManagement2023} and SGLang~\cite{zhengSGLangEfficientExecution}.
Training is dominated by dense forward/backward computation, gradient synchronization, and optimizer-state management, and is commonly executed with systems such as FSDP~\cite{zhaoPyTorchFSDPExperiences2023} and Megatron-LM~\cite{shoeybiMegatronLMTrainingMultiBillion2020}.
BiDiRL is complementary to these engines: it treats them as stage executors and optimizes cross-stage resource planning and runtime scheduling above them.

\textbf{RL frameworks for LLMs.}
Existing LLM RL frameworks make different resource-efficiency trade-offs.
NeMo-Aligner~\cite{shenNeMoAlignerScalableToolkit2024a} and OpenRLHF~\cite{huOpenRLHFEasytouseScalable} adopt disaggregated execution, which simplifies integration of heterogeneous rollout and training engines but can leave one stage idle while the other is active.
Colocated systems such as veRL~\cite{shengHybridFlowFlexibleEfficient2025} and RLHFuse~\cite{zhongOptimizingRLHFTraining} improve utilization by time-sharing the same GPU pool across RL stages.
However, colocation also couples rollout and training to one resource pool and one set of parallelism constraints, despite their different scaling behavior.
StreamRL~\cite{zhongStreamRLScalableHeterogeneous} and AReaL~\cite{fuAReaLLargeScaleAsynchronous2025} reduce pipeline bubbles through asynchronous execution, but still keep committed rollout and trainer pools fixed within an execution window.
Elastic systems such as StreamRL and RLBoost~\cite{wuRLBoostHarvestingPreemptible2025} further accelerate rollout with idle or external resources, but they primarily expand the rollout side rather than support bidirectional borrowing between committed rollout and trainer pools.

In contrast, BiDiRL targets bidirectional borrowing under a fixed committed budget: idle rollout resources can execute trainer work, and idle trainer resources can execute rollout work when the predicted benefit exceeds switching overhead.

\section{Conclusion}
\label{sec:conclusion}

This paper presents BiDiRL, a bidirectional resource scheduler for asynchronous and disaggregated LLM RL.
BiDiRL treats resource bubbles as a two-timescale scheduling problem.
Before execution, a static planner reduces structural bubbles by selecting a throughput-aware, hot-switch-compatible resource envelope.
During execution, a model-guided scheduler recovers residual bubbles by lending idle resources from either committed pool to the current bottleneck stage when the predicted gain covers switching cost.
The hot-switch runtime executes admitted borrowing, while online profiling keeps stage-time models and switching costs calibrated.
Across workloads, staleness settings, model sizes, datasets, and hardware platforms, BiDiRL improves throughput over existing systems by up to \(1.94\times\) while preserving the logical RL dataflow and convergence behavior.

\bibliographystyle{acm_reference_format}
\bibliography{refs}

\appendix
% !TEX root = ../paper.tex

\section{Appendix}
\label{app:artifact}

\subsection{Stage-Time Models}
\label{app:planner-models}

This appendix gives the stage-time models exposed to the static planner and runtime scheduler.
The main text uses only their prediction interfaces:
\(\mathcal{M}_r(n,d)\) predicts the time for \(d\) rollout replicas to produce \(n\) prompt-level rollout groups, and
\(\mathcal{M}_t(U,d)\) predicts the time for \(d\) trainer replicas to consume a stage-ordered trainer workload \(U\).
Both models use replica-max semantics.
Given a workload \(X\) and \(d\) replicas, the workload is partitioned into replica-local workloads \(X_1,\ldots,X_d\), and the stage time is the slowest replica:
\begin{equation}
\mathcal{M}_{\star}(X,d)=\max_{j\in[1,d]}T_{\star,\mathrm{rep}}(X_j),
\qquad \bigcup_{j=1}^{d} X_j=X .
\end{equation}
This form captures the synchronization point at the end of a rollout or trainer stage while keeping model evaluation lightweight.

\subsubsection{Trainer Model}
\label{app:trainer-model}

The trainer workload consists of ordered micro-batches for old-log-probability computation, reference-log-probability computation, and actor update.
Let \(s\in\mathcal{S}\) denote one such trainer stage.
For a micro-batch \(b\), each sample \(u\in b\) has prompt length \(p_u\) and response length \(r_u\).
The forward FLOPs are
\begin{equation}
    \begin{aligned}
    F(b)
    &= 2N_{\mathrm{nonemb}}\sum_{u\in b}(p_u+r_u)\\
    &\quad +4LH\sum_{u\in b}(p_u+r_u)^2\\
    &\quad +2HV\sum_{u\in b}r_u .
    \end{aligned}
\end{equation}
The multiplier \(m_s\) is \(1\) for old/ref forward stages and \(3\) for actor update.
The communication term is fitted with an \(\alpha\)-\(\beta\) model over the stage-specific message size \(n_s(b)\),
\begin{equation}
C_s(b)=\alpha_{\mathrm{comm},s}+\beta_{\mathrm{comm},s}n_s(b) .
\end{equation}
Compute and communication are combined with a fitted \(p\)-norm overlap~\cite{qiaoPolluxCoadaptiveCluster},
\begin{equation}
\mathrm{overlap}(x,c,r)=(x^r+c^r)^{1/r}.
\end{equation}
The predicted micro-batch time is
\begin{equation}
t_s(b)=\tau_s+
\mathrm{overlap}(\alpha_s m_s F(b), C_s(b), r_s),
\end{equation}
where \(\tau_s\), \(\alpha_s\), and \(r_s\) are calibrated from profiling runs.
For replica \(j\), the trainer replica time sums the ordered micro-batch times assigned to that replica:
\begin{equation}
T_{t,\mathrm{rep}}(U_j)=\sum_{s\in\mathcal{S}}\sum_{b\in U_{j,s}} t_s(b).
\end{equation}
The trainer stage-time model \(\mathcal{M}_t(U,d)\) then follows the replica-max rule above.

\subsubsection{Rollout Model}
\label{app:rollout-model}

For a rollout request set \(R_j\) assigned to one rollout replica, request \(i\in R_j\) has prompt length \(p_i\) and response length \(r_i\).
The per-token KV-cache footprint is
\begin{equation}
b_{\mathrm{kv}}=\frac{2LH_{\mathrm{kv}}d_{\mathrm{head}}b_{\mathrm{dtype}}}{tp},
\end{equation}
where \(tp\) is the tensor-parallel degree.
At average live length \(\ell\), the KV-limited concurrency is
\begin{equation}
N_{\max}(\ell)=\max\left(1,\left\lfloor\frac{B_{\mathrm{kv}}}{b_{\mathrm{kv}}\ell}\right\rfloor\right).
\end{equation}
For prefill, with \(B_j=|R_j|\),
\begin{equation}
N_{\mathrm{pre}}=\min(B_j,N_{\max}(\mathrm{mean}(p_i))),
\qquad
W_{\mathrm{pre}}=\left\lceil\frac{B_j}{N_{\mathrm{pre}}}\right\rceil,
\end{equation}
\begin{equation}
F_{\mathrm{pre}}=2N_{\mathrm{nonemb}}\sum_{i\in R_j}p_i
+4LH\sum_{i\in R_j}p_i^2+2HVB_j,
\end{equation}
\begin{equation}
P_{\mathrm{pre}}=\tau_{\mathrm{pre}}W_{\mathrm{pre}}+
\alpha_{\mathrm{pre}}F_{\mathrm{pre}} .
\end{equation}
For decode step \(t\), let \(A_t=\{i\in R_j\mid r_i\ge t\}\) and \(\ell_t=\mathrm{mean}_{i\in A_t}(p_i+t-1)\).
The number of decode waves is
\begin{equation}
w_t=\left\lceil\frac{|A_t|}{\min(|A_t|,N_{\max}(\ell_t))}\right\rceil .
\end{equation}
The decode features are
\begin{equation}
    \begin{aligned}
    W_{\mathrm{dec}} &= \sum_t w_t,\\
    N_{\mathrm{tok}} &= \sum_i r_i,\\
    H_{\mathrm{dec}}
        &= \sum_i\left(r_i p_i+\frac{r_i(r_i-1)}{2}\right).
    \end{aligned}
\end{equation}
The decode predictor is
\begin{equation}
P_{\mathrm{dec}}=\tau_{\mathrm{dec}}W_{\mathrm{dec}}
+\beta_{\mathrm{tok}}N_{\mathrm{tok}}+\beta_{\mathrm{hist}}H_{\mathrm{dec}} .
\end{equation}
The rollout replica time is
\begin{equation}
T_{r,\mathrm{rep}}(R_j)=
\mathrm{overlap}(P_{\mathrm{pre}},C_{\mathrm{pre}},r)+
\mathrm{overlap}(P_{\mathrm{dec}},C_{\mathrm{dec}},r).
\end{equation}
Given \(n\) prompt-level rollout groups and \(d\) rollout replicas, \(\mathcal{M}_r(n,d)\) partitions the corresponding request set across replicas and applies the replica-max rule.

\end{document}